\documentclass[11pt]{article} 
\usepackage{amsfonts,amscd,amsmath,amssymb,graphicx}

\newcommand{\oh}{\frac{1}{2}}
\def\4{\tfrac{1}{4}}
\def\ep{\text{e}}

\def\g{\mathfrak{g}}

\def\z{z_{\text{\tiny 0}}}
\def\zt{z_{\text{\tiny T}}}

\def\T{T_{\sigma_s}}
\def\TD{T_1}
\def\zz{z_{\text{\tiny min}}}
\def\zm{z_{\text{\tiny max}}}

\def\sig{\sigma_{\text{\tiny 0}}}
\def\I{\mathfrak{c}}
\def\sT{\sigma_{\text{\tiny T}}}
\def\F{F}

\def\ri{r^{(1)}}
\def\rii{r^{(2)}}
\def\riim{r^{(2)}_{\text{\tiny max}}}
\def\rmax{r_{\text{\tiny max}}}
\title{On Heavy-Quark Free Energies, Entropies, Polyakov Loop, and AdS/QCD}
\author{Oleg Andreev\thanks{andre@itp.ac.ru}\\\\
{\it Max-Planck Institut f\" ur Physik, F\" ohringer Ring 6, 80805 M\" unchen, Germany;}\\
{\it L.D. Landau Institute for Theoretical Physics, Kosygina 2, 119334 Moscow, Russia}
\\\\
Valentin I. Zakharov\thanks{xxz@mppmu.mpg.de}\\\\
{\it Istituto Nazionale di Fisica Nucleare -- Sezione di Pisa} \\
{\it Dipartimento di Fisica Universita di Pisa, Largo Pontecorvo 3, 56127 Pisa, Italy;}\\
{\it Max-Planck Institut f\" ur Physik, F\" ohringer Ring 6, 80805 M\" unchen, Germany} }
\date{}

\begin{document} 

\vspace{-8cm} 
\maketitle 
\begin{abstract} 
In this paper we explore some of the features of a heavy quark-antiquark pair at finite temperature using a five-dimensional framework 
nowadays known as AdS/QCD. We shall show that the resulting behavior is consistent with our qualitative expectations of thermal gauge theory. 
Some of the results are in good agreement with the lattice data that provides additional evidence for the validity of the proposed model. \\
PACS: 12.38.Lg, 12.39.Pn, 12.90.+b
 \end{abstract}

\vspace{-14cm}
\begin{flushright}
MPP-2006-162
\end{flushright}
\vspace{12.5cm}

%__________________________  I N T R O ______________________________
\section{Introduction}
 \renewcommand{\theequation}{1.\arabic{equation}}
\setcounter{equation}{0}

This is our second paper of a series devoted to the thermal properties of a pure gauge theory within a five-dimensional framework 
nowadays known as AdS/QCD. We explore further the model proposed in the first paper \cite{az2}. It is based on the following Euclidean 
background metric

\begin{equation}\label{metric}
ds^2
=R^2\frac{h}{z^2}
\left(f dt^2+d\vec{x}^2+\frac{1}{f}dz^2\right)
\,,\quad\quad 
h(z)=\ep^{\oh cz^2}
\,,\quad\quad
f(z)=1-\bigl(\tfrac{z}{\zt}\bigr)^4
\,,
\end{equation}
where $t$ is a periodic variable of period $\pi\zt$ such that $\zt=\tfrac{1}{\pi T}$, with $T$, the temperature. At zero temperature, we have 
in fact the slightly deformed $\text{AdS}_5$ metric. Such a deformation is notable because it results in a Regge-like spectrum. This fact allows 
one to fix the value of $c$ from the $\rho$ meson trajectory. For our purposes, we use the estimate of  \cite{oa}

\begin{equation}\label{c}
c\approx 0.9\,\text{GeV\,}^2
\,.
\end{equation}

It is worth noting that the metric \eqref{metric} {\it does not} contain any free fit parameter. Moreover, the only dimensionful parameter of our 
model is the Regge slope of meson trajectories. It is quite unusual for QCD, where it is $\Lambda_{\text{\tiny QCD }}$. Thus, evaluations of 
thermodynamic quantities and the Polyakov loop we are going to undertake can be considered as a further consistency check of our model.

For the case of interest, let us briefly point out a couple of facts. 

\noindent (i) The free energy of a heavy (static) quark-antiquark pair at finite temperature is expressed in terms of a correlator of two 
Polyakov loops \cite{svet}

\begin{equation}\label{free-e}
\F (r,T)=-T\ln
\langle
L(\vec{x}_1)L^\dagger (\vec{x}_2)
\rangle 
+Tc(T)
\,,
\end{equation}
with $r=\lvert\vec{ x}_1-\vec{x}_2\rvert$ and $\vec{x}_i$ being a point in $R^3$. In \eqref{free-e} the free energy is defined 
up to a normalization constant $c(T)$ which is related to the infinite self-energy of the quarks.

There is a subtle point here. In the literature $F$ is often called the heavy quark potential at finite temperature. Apparently, such a definition 
discards the entropy contribution.\footnote{For a discussion of this issue, see, e.g., \cite{karsch2} and references therein.} 

An order parameter for the confinement-deconfinement phase transition is the expectation value of the Polyakov loop. After the normalization 
of \eqref{free-e}, it is then

\begin{equation}\label{L}
L=\exp\bigl\{-\tfrac{1}{2T}F(r=\infty,T)\bigr\}
\,.
\end{equation}

\noindent (ii) In discussing a Wilson line within AdS/CFT (QCD) \cite{Tloops}, one first chooses a contour ${\cal C}$ on a four-manifold which is 
the boundary of a five-dimensional manifold. Next, one has to study fundamental strings on this manifold such that the string world-sheet 
has ${\cal C}$ for its boundary. The expectation value of the loop is schematically given by the world-sheet path integral 

\begin{equation}\label{wilson}
\langle\,W({\cal C})\,\rangle=\int DX\,\ep^{-S_w}
\,,
\end{equation}
where $X$ denotes a set of world-sheet fields. $S_w$ is a world-sheet action.

In principle, the integral can be evaluated semiclassically in terms of minimal surfaces that obey the boundary conditions. The result is 
written as

\begin{equation}\label{wilson2}
\langle\,W({\cal C})\,\rangle=\sum_n w_n\ep^{-S_n}
\,,
\end{equation}
where $S_n$ means a regularized minimal area whose relative weight is $w_n$.\footnote{The point is that the areas are divergent but the 
divergences are proportional to the circumference of ${\cal C}$.}

The paper is organized as follows. In section 2, we develop the framework we will work. In section 3, we present, on the basis of AdS/QCD, 
a few results on a heavy quark-antiquark pair in a thermal medium. In particular, we exhibit the free energy, the string tension, and 
the entropy at low temperatures as well as the Polyakov loop expectation value. We conclude in section 4 with a discussion of some open 
problems. 

%____________________________________________________________________________________________
\section{General Formalism}
\renewcommand{\theequation}{2.\arabic{equation}}
\setcounter{equation}{0}

Our basic approach will be as follows. It is believed that the correlator of the two Polyakov loops has a path integral representation like that of the 
Wilson loop.\footnote{Note that the corresponding world-sheet has now two boundaries.} Given the background metric, we can attempt to 
evaluate the values of the regularized areas $S_n$. If we then discard quantum fluctuations of strings, in the case of one dominant exponent
we will get that the free energy is simply proportional to a proper $S_\ast$

\begin{equation}\label{approx}
F=TS_\ast 
\,.
\end{equation}
Note that the regularized areas are defined up to normalization constants. Since these constants are in fact due to the infinite self-energy of 
the quark sources, we omit $c(T)$. Moreover, $w_\ast$ can be absorbed into a proper normalization constant too. In this paper we will 
use the approximation \eqref{approx}. 

But before going on, let us shortly pause here to gain some intuition about the problem at hand. As in \cite{az2},  we introduce 
the notion of an effective string tension depending on the fifth coordinate $z$. It is given by\footnote{It follows from Eq.\eqref{ng}.}

\begin{equation}\label{eff}
\sigma (z)=\frac{h}{z^2}\sqrt{f}(z)
\,.
\end{equation}

Now consider the behavior of a string bit in the effective potential $V=\sigma (z)$. An important observation is that 
the form of $V$ is temperature dependent. Indeed, a short algebra shows that there is a special value of temperature $\TD=\frac{1}{\pi}
\sqrt{\frac{c}{\sqrt{27}}}\approx 130\,\text{MeV}$.\footnote{We use \eqref{c} for all estimates. } Below $\TD$ the effective potential has local 
extrema at 

\begin{equation}\label{zmin}
\zz=\zt\sqrt{\tfrac{2}{\sqrt{3}}
\sin\Bigl(\tfrac{1}{3}
\arcsin\tfrac{T^2}{\TD^2}
\Bigr)}
\,,\quad\quad
\zm=\zt\sqrt{\tfrac{2}{\sqrt{3}}
\sin\Bigl(\tfrac{\pi}{3}-
\tfrac{1}{3}
\arcsin\tfrac{T^2}{\TD^2}
\Bigr)}\,,
\end{equation}
while above this temperature the potential is just a decreasing function of $z$ as shown in Fig.1. 
%________________________  f - 1  __________________________________
%
\vspace{.6cm}
\begin{figure}[htbp]%[htbp]
\centering
\includegraphics[width=5.5cm]{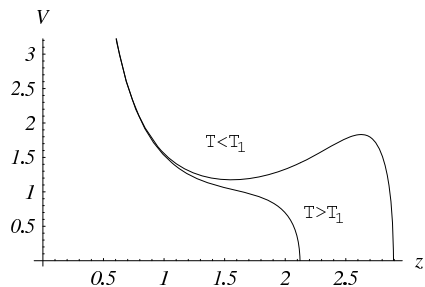}
\caption{\small{Schematic representation of the effective potential below and above $\TD$.}}
%\label{fig:graph1}
\end{figure}
%______________________________________________________________

Such a behavior clearly has some of the suspected properties of gauge theory at finite temperature.\footnote{The notion of 
the effective potential turned out to be useful in studying symmetry breaking (phase transition) within field theories at finite 
temperature \cite{effpot}. Like in field theory, our model also has the effective potential whose form is temperature dependent. } For 
temperatures below $\TD$ the string ended on the heavy quark-antiquark pair set at $z=0$ can not get deeper than $\zz$ in $z$ direction 
because a repulsive force prevents it from doing so. This gives rise somewhat of a wall located at $z=\zz$. The large distance physics of the 
string is determined by this wall. Since the value of the effective potential (string tension) at its minimum is not vanishing, the quark-antiquark 
free energy has a dominant linear term whose coefficient is proportional to $\sigma (\zz )$. So, this can indeed be interpreted as the low 
temperature phase. On the other hand, for temperatures above $\TD$ the string can get deeper and finally reach the horizon $z=\zt$. 
The large distance physics is now determined by the near horizon geometry. The crucial point is that the effective potential vanishes on the 
horizon. As a result, there is no linear term in the quark-antiquark free energy. This can be interpreted as the deconfined 
phase.\footnote{Although the above arguments indicate that the phase transition occurs, we should caution the reader that it is a qualitative 
way of thinking about the problem at hand. }

Now let us go back to the approach and choose a pair of the contours (Polyakov loops) living on the boundary ($z=0$) of our five dimensional 
space. We set 

\begin{equation}\label{loops}
\vec{x}_1=(-\tfrac{r}{2},0,0)
\,,\quad\quad
\vec{x}_2=(\tfrac{r}{2},0,0)
\,.
\end{equation}

Next we want to look for static configurations of the world-sheet action. To this end, we make use of the Nambu-Goto action equipped with the 
background metric \eqref{metric}

\begin{equation}\label{ng0}
S=\frac{1}{2\pi\alpha'}\int d^2\xi\,\sqrt{\det \, G_{nm}^{}\partial_\alpha X^n\partial_\beta X^m\vphantom{\bigl(\bigr)}}
\,.
\end{equation}
There are basically two types of configurations to be considered. One type describes connected surfaces whose boundaries are the Polyakov 
loops, while another describes disconnected surfaces. 

Our first goal will be to analyze connected surfaces. In this case the world-sheet 
coordinates $\xi$'s can be chosen as $\xi_1=t$ and $\xi_2=x$. With such a choice, the action takes the form

\begin{equation}\label{ng}
S_w=\frac{\g }{2\pi T}\int^{\,\tfrac{r}{2}}_{-\tfrac{r}{2}} 
dx\,\frac{h}{z^2}
\sqrt{f+(z')^2}
\,,
\end{equation}
where $\g=\tfrac{R^2}{\alpha'}$. A prime denotes a derivative with respect to $x$.

It is easy to find the equation of motion for $z$

\begin{equation}\label{eqm}
zz''+\left(f+(z')^2\right)\left(2-z\partial_z\ln h\right)-
\Bigl(\tfrac{1}{2} f+(z')^2\Bigr)z\partial_z\ln f=0
\end{equation}
as well as its first integral
\begin{equation}\label{int}
\I=\frac{hf}{z^2\sqrt{f+(z')^2}}
\,.
\end{equation}

On symmetry grounds, we have $z'\vert_{x=0}=0$. This allows us to express the integration constant $\I$ via the value of 
$z$ at $x=0$. So, we get

\begin{equation}\label{I}
\I=\sigma\vert_{z=\z}
\,\,,
\end{equation}
where $\sigma$ is defined by \eqref{eff} and $\z=z\vert_{x=0}$.

Next we perform the integral over $\left[-\tfrac{r}{2},\tfrac{r}{2}\right]$ of $dx$. By virtue of \eqref{int}, it is given by 

\begin{equation}\label{x}
r=2\int_{\cal C} \frac{dz}{\sqrt{f}}\left(\Bigl(\frac{\sigma}{\I}\Bigr)^2-1\right)^{-\oh}
\,,
\end{equation}
where ${\cal C}$  is a contour in $z$ plane.

We look for solutions that obey the following condition $\z=\max z$. The reason for this is that, on general grounds, the string ended on 
the quarks set at $x=\pm r/2$ reaches the deepest point in $z$ direction at $x=0$. 

For temperatures below $\TD$, there are two possibilities for $\cal C$ and, as a result, we have

\begin{align}
\ri &=2\int_0^{\z}\frac{dz}{\sqrt{f}}\left(\Bigl(\frac{\sigma}{\I}\Bigr)^2-1\right)^{-\oh}
\,\,\,\,,
\qquad
\text{with}
\quad 
0 \leq\z\leq\zz
\,\,\,\,,
\label{r1}\\
\rii &=2\int_{\zm}^{\z}\frac{dz}{\sqrt{f}}\left(\Bigl(\frac{\sigma}{\I}\Bigr)^2-1\right)^{-\oh}
\,,
\qquad
\text{with}
\quad
\zm\leq\z\leq\zt
\,,\label{r2}
\end{align}
but otherwise $r$ is complex. After a short inspection we find that $\ri$ is a continuously growing function of $\z$ on the interval 
$[0,\zz]$. Moreover, it equals to zero at $\z=0$ and goes to infinity as $\z\rightarrow\zz$. The function $\rii$ is, unlike $\ri$, decreasing.
It goes from its maximum (finite) value $\riim$ at $\z=\zm$ to zero at $\z=\zt$ (on the horizon). Thus, the second solution contributes only 
at distances smaller than $\riim$. To complete the picture, we present the plots of $r^{(i)}$ in Fig.2.
%________________________  f - 2  __________________________________
%
\vspace{.6cm}
\begin{figure}[ht]%[htbp]
\centering
\includegraphics[width=5.5cm]{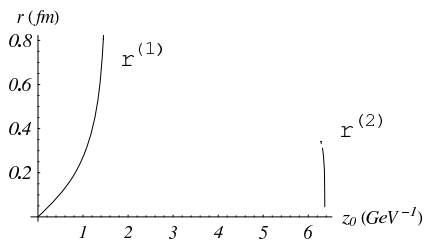}
\caption{\small{Typical graphs of $r^{(i)}$ below $\TD$. Here $T=0.05\,\text{GeV}.$}}
%\label{fig:graph1}
\end{figure}
%______________________________________________________________
%\vspace{-0.3cm}

When temperature is increased, the interval $\left[\zz ,\zm \right]$ becomes smaller and finally disappears at $T=\TD$. For $T>\TD$, a simple 
analysis leads to the picture that differs noticeably from that of Fig.2. The point is that both the solutions now contribute only at distances smaller 
than some finite $\rmax$. This is illustrated in Fig.3.

Now we move on to the second type that describes disconnected surfaces. In this case a surface contains two pieces each of which has a 
topology of a cylinder. The cylinders are stretched from the Polyakov loops on the boundary to the horizon. If we use $\xi_1=t$ and $\xi_2=z$ as 
the world-sheet coordinates, the Nambu-Goto action is then
%________________________  f - 3 __________________________________
%
\vspace{.6cm}
\begin{figure}[ht]%[htbp]
\centering
\includegraphics[width=5.5cm]{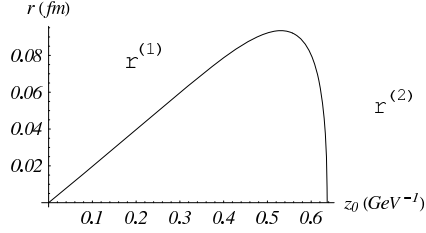}
\caption{\small{Typical graphs of $r^{(i)}$ above $\TD$: $r^{(1)}$ is growing from $0$ to $\rmax$, 
while $r^{(2)}$ is decreasing from $\rmax$ to $0$. We set $T=0.5\,\text{GeV}$.}}
%\label{fig:graph1}
\end{figure}
%______________________________________________________________
%\vspace{-0.3cm} 

\vspace{0.3cm}
\begin{equation}\label{2ng}
S_w=\frac{\g }{2\pi T}\int^{\zt}_0 
dz\,\frac{h}{z^2}
\sqrt{1+f (\dot{\vec{x}})^2}
\,,
\end{equation}
where a dot stands for a derivative with respect to $z$. The equation of motion for $x$ is

\begin{equation}\label{eq-x}
\frac{d}{d z}\biggl(\sigma\dot{\vec{x}}\Bigl(f^{-1}+(\dot{\vec{x}})^2\Bigr)^{-\oh}\biggr)=0
\,.
\end{equation}
It is obvious that it has a trivial solution $\vec{x}=const$ that represents a straight string stretched between the boundary and the horizon. We will 
call the choice \eqref{loops} the solution $r^{(\infty )}$. Since this solution makes the dominant contribution, as seen from the integrand 
in \eqref{2ng}, we will not dwell on other solutions here. 

Having discussed the solutions, we  can now see what happens with the corresponding areas. We begin with the solution $r^{(1)}$. To this end, we 
use the first integral to reduce the integral over $x$ to that over $z$ in \eqref{ng}. Since the integral is divergent at $z=0$ due to the factor $z^{-2}$ 
in the background metric, in the process we regularize it by imposing a cutoff $\epsilon$. At the end of the day, we have

\begin{equation}
S^{\text{\tiny R}}_1=\frac{\g}{\pi T}\int_ \epsilon^{\z}\frac{dz}{z^2}
h\biggl(1-\Bigl(\frac{\I}{\sigma}\Bigr)^2\biggr)^{-\oh}
\,.
\end{equation}
Subtracting the $\frac{1}{\epsilon}$ term we find a finite result

\begin{equation}\label{F1}
S_1=-\frac{\g}{\pi T\z}+\frac{\g}{\pi T}\int_ 0^{\z}\frac{dz}{z^2}
\biggl[h\biggl(1-\Bigl(\frac{\I}{\sigma}\Bigr)^2\biggr)^{-\oh}
-1
\biggr]+\frac{c(T)}{T}
\,,
\end{equation}
where $c(T)$ stands for a normalization constant. 

For the second solution $r^{(2)}$, it is a little bit tricky because the minimal surface is built by sewing together two pieces. The first piece comes 
from $r^{(\infty)}$ defined on the interval $[0,\zm ]$. At $z=\zm$, it is sewn with the second piece coming from $r^{(2)}$. The integral 
\eqref{2ng} is divergent at $z=0$, so we regularize it by imposing the cutoff $\epsilon$ as before. After subtracting the divergency, we get

 \begin{equation}\label{F2}
S_2=-\frac{\g}{\pi T\zm}+\frac{\g}{\pi T}\int^{\zm}_ 0 \frac{dz}{z^2}\left( h-1\right)+
\frac{\g}{\pi T}\int_ {\zm}^{\z}\frac{dz}{z^2}
h\biggl(1-\Bigl(\frac{\I}{\sigma}\Bigr)^2\biggr)^{-\oh}
+\frac{c(T)}{T}
\,.
\end{equation}
Note that both the areas are regularized in the same way, so the corresponding normalization constants coincide.

For the third solution, the minimal area can be read off from Eq.\eqref{F2}. One drops the third term as coming from the solution $r^{(2)}$ and 
replaces $\zm$ with $\zt$ in the remaining terms. The area is then

\begin{equation}\label{S3}
S_\infty=-\g+\frac{\g}{\pi T}\int^{\zt}_ 0 \frac{dz}{z^2}\left( h-1\right)
+\frac{c(T)}{T}
\,.
\end{equation}

Now let us look at the $S_i$ 's as functions of $r$. We begin with sufficiently low temperature. Since the solution $r^{(2)}$ is defined only for 
distances smaller than $\riim$, it makes sense to first probe asymptotic behaviors near the point $r=0$. A short inspection shows that $S_1$ is 
not bounded from below, while the others are bounded. This provides a bit of evidence in favor of dominance of $S_1$ at small distances. Further 
numerical calculations show that this is indeed the case. Moreover, the first solution turns out to be dominant at physical 
distances too.\footnote{We mean the interval $0.2\,\text{fm}\lesssim r\lesssim 2\,\text{fm}$ that is of primary importance for phenomenology.}
We present the plot of the regularized areas in Fig.4.\footnote{The overall constant $\g$ can be fixed from the slope of the heavy quark potential 
at zero temperature. We use the estimate of \cite{az} $\g\approx 0.94$, here and below.}
%________________________  f - 4 __________________________________
%
\vspace{.6cm}
\begin{figure}[htbp]
\centering
\includegraphics[width=5.5cm]{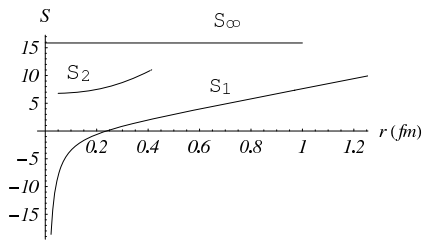}
\caption{\small{Typical graphs of $S_i$. Here $T=0.1\,\text{GeV}$ and $c(T)=0$.}}
\end{figure}
%______________________________________________________________
%\vspace{-0.3cm} 

From this figure it is also clear that there exists a critical distance $r_c$ such that for larger distances $S_\infty$ becomes dominant. There is 
an apparent reason for this. At very large distances the quark and the anti-quark decouple from each other. Usually, instability occurs in 
models with dynamical quarks, where a string breaks. In the case of interest this occurs due to emission of closed strings (glueballs). 
As known, in AdS/QCD a free particle is described by a straight string stretched between the boundary and the horizon that is nothing but 
our  solution $r^{(\infty)}$.

When temperature is increased,  two effects are seen: (1) The minimal difference between the functions $S_1$ and $S_2$  becomes smaller. (2) 
The critical distance $r_c$ is decreasing. For temperatures close to $T=\TD$, $S_1$ is no longer dominant at physical distances 
and it is time to account for the other solutions in the series \eqref{wilson2}. 

%_______________________________________________________________________________________

\section{Applications}
\renewcommand{\theequation}{3.\arabic{equation}}
\setcounter{equation}{0}
In this section we will consider some applications of the developed formalism. We will focus on the cases where the approximation \eqref{approx} 
might be applicable.

\subsection{Free Energy at Low Temperature}

We begin with temperatures which are sufficiently smaller than $\TD$. In this case the free energy can be evaluated by using \eqref{approx} 
with $S_\ast=S_1$.\footnote{In the remaining part of this section we omit the index specifying the solution each time when the meaning is clear 
from the context.}

First, we need to fix the normalization constant $c(T)$. In doing so, we follow \cite{karsch} and look for the small $r$ expansion of the free energy. 
Since small distances correspond to small deviations in $z$ direction, we need to study the expressions \eqref{r1} and 
\eqref{F1} near $\z=0$. The asymptotic behavior of $r(\z)$ is given by\footnote{To this order, the calculation is identical to that of \cite{az}.}

\begin{equation}\label{small-r}
r=\frac{1}{\rho}\z -\frac{c}{4\rho} (1-\pi\rho^2)\z^3+O(\z^5)
\,,
\end{equation}
where $\rho=\Gamma^2\bigl(\tfrac{1}{4}\bigr)/(2\pi)^{\tfrac{3}{2}}$. 

In a similar way we find the behavior of $F$. It is 

\begin{equation}\label{small-F}
F=-\frac{\g}{2\pi\rho}\frac{1}{\z}+c(T)+\frac{\g c}{8\pi\rho}(3\pi\rho^2-1)\z+O(\z^3)
\,.
\end{equation}
Combining this with \eqref{small-r}, we get 

\begin{equation}\label{small-F1}
F=-\frac{\kappa_0}{r}+c(T)+\sigma_0r+O(r^3)
\,,
\end{equation}
where $\kappa_0=\frac{\g}{2\pi}\rho^{-2}$ and $\sigma_0=\frac{\g}{4} c\rho^2$.

Having derived the small distance expansion of the free energy, we can now get that of the singlet free energy\footnote{Here we assume a pure 
$SU(3)$ gauge theory.}

\begin{equation}\label{singletF}
F_1=F-T\ln 9=-\frac{\kappa_0}{r}+c(T)-T\ln 9+O(r)
\,
\end{equation}
and compare it with the small distance expansion of the heavy-quark potential at zero temperature. To leading order, we have the same Coulomb 
term as in \cite{az}. To get an agreement at next-to-leading order, we choose 

\begin{equation}\label{cT}
c(T)=T\ln 9 +C
\,.
\end{equation}
Finally, we fix the value of $C$ by matching it to the constant term of the Cornell potential \cite{cornell}. This yields 

\begin{equation}\label{C}
C=-0.25\,\text{GeV}
\,.
\end{equation}

Actually, our result \eqref{singletF} shows that at sufficiently small distances the singlet free energy of the pair is 
temperature independent. So, the agreement with the lattice data \cite{karsch2,karsch} is very satisfactory at this point.

Our next goal is to analyze the long distance behavior of $F$. As noted in section 2, large $r$ corresponds to $\z\sim\zz$. So, we need to study 
the behavior of \eqref{r1} and \eqref{F1} near $\z=\zz$. In this case a crucial observation is that the 
integrals are dominated by the upper limits, where they take the form $\int^1 dv/\sqrt{a(1-v)+b(1-v)^2}$. Such an integral may be 
found in \cite{gr}. At the end of the day, we have 

\begin{equation}\label{r-large}
r=-w\ln\left(\zz-\z\right)+O(1)
\,, \quad\quad
\F=-\sT w\ln\left(\zz-\z\right)+O(1)
\,,
\end{equation}
where $w=2\sqrt{\sigma/(f\sigma'')}\bigl\vert_{z=\zz}$ and $\sT=\frac{\g}{2\pi}\sigma\bigl\vert_{z=\zz}$.

This means that at long distances the free energy of the pair shows the desired confining behavior

\begin{equation}\label{F-large1}
\F=\sT r + O(1)
\,,
\end{equation}
with the string tension 
\begin{equation}\label{sigma-T}
\sT=\sigma\,\frac{\ep^{\,t-1}}{t}
\left(1-t^2\,\frac{T^4}{\T^4}\right)^{\oh}
\,,\quad\quad
t=3\,\frac{\TD^2}{T^2}\sin\Bigl(\tfrac{1}{3}\arcsin\tfrac{T^2}{\TD^2}\Bigr)
\,.
\end{equation}
Here $\sigma$ denotes a tension at zero temperature. Explicitly, it is given by $\sigma=\tfrac{\g\,\ep}{4\pi}c$ \cite{az}. We have 
also introduced the critical temperature $\T=\frac{1}{\pi}\sqrt{\frac{c}{2}}\approx 210\,\, \text{MeV}$ obtained from the spatial string 
tension in \cite{az2}.

As expected, the string tension $\sT$ is a decreasing function of T. In Fig.5 we have plotted $\sT /\sigma$ against 
$T/\TD$. 
%________________________  f - 5 __________________________________
%
\vspace{.6cm}
\begin{figure}[ht]%[htbp]
\centering
\includegraphics[width=5.5cm]{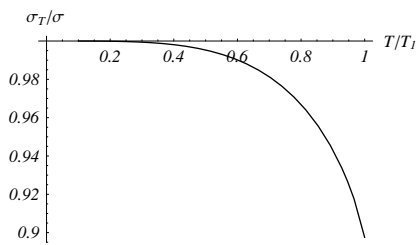}
\caption{\small{String tension in units of $\sigma$ versus temperature in units of $T_1$.}}
%\label{fig:graph1}
\end{figure}
%______________________________________________________________

We conclude this subsection by making a few remarks.
\newline (i) It is quite interesting that the tension shows {\it very} little dependence on temperature up to 
$T\approx 0.8\TD\approx 100\,\,\text{MeV}$. 
\newline (ii) Unlike the coefficient $\sT$ of the linear term in the large distance expansion of $F$, the coefficient $\sig$ of the 
linear term in the small distance expansion turns out to be independent of temperature.
\newline (iii) The free energy of the pair is written in parametric form given by Eqs.\eqref{r1} and \eqref{F1}. Since 
we do not know how to eliminate the parameter $\z$ and find $F$ as a function of $r$, we present the result of numerical 
calculations. In Fig.6 the free energy $F$ versus $r$ for a temperature below $\TD$ is shown.

%________________________  f - 6__________________________________
%
\vspace{.6cm}
\begin{figure}[ht]%[htbp]
\centering
\includegraphics[width=5.5cm]{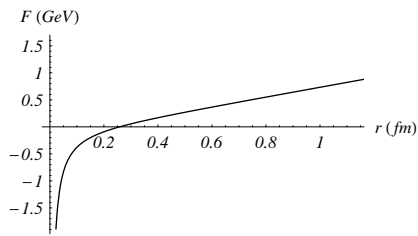}
\caption{\small{Free energy versus $r$ at $T=0.1\,\text{GeV}$.}}
%\label{fig:graph1}
\end{figure}
%______________________________________________________________

%_______________________________________________________________________________________
\subsection{Entropy and Energy at Low Temperature}

We have already mentioned that in the literature the entropy contributions are sometimes ignored.\footnote{This is the case for 
AdS/CFT(QCD) too. See, e.g., \cite{Tloops}.} It is therefore of great importance to address this issue. 

Having understood the behavior of the free energy, we can easily investigate the properties of the entropy. Since the free energy 
shows no temperature dependence at small distances, it makes sense to focus on long distances, where its 
behavior is given by \eqref{F-large1}. In this case a straightforward calculation leads to the following expression

\begin{equation}\label{ent}
{\cal S}=-\left(\frac{\partial F}{\partial T}\right)_r
=-\frac{1}{2}\sT r \left(\frac{\partial\ln f}{\partial T}\right)_{\zz}
\,.
\end{equation}
Here we treat $f$ as a function of two variables. Then, using the relations \eqref{metric} and \eqref{zmin}, we can compute the entropy 
density of the pair

\begin{equation}\label{ent-d}
\frac{{\cal S}}{r}=8\frac{\sT}{T}
\frac{\sin^2\Bigl(\tfrac{1}{3}\arcsin\tfrac{T^2}{\TD^2}\Bigr)}{3-4\sin^2\Bigl(\tfrac{1}{3}\arcsin\tfrac{T^2}{\TD^2}\Bigr)}
\,.
\end{equation}

A closer look at this expression shows that the entropy density grows as the temperature increases. This is the expected result. To complete the 
picture, we plot the entropy density against $T/\TD$ in Fig. 7.
%________________________  f - 7__________________________________
%
\vspace{.6cm}
\begin{figure}[hbtp]
\centering
\includegraphics[width=5.5cm]{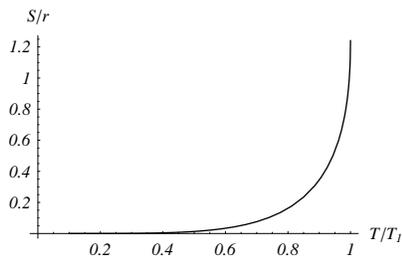}
\caption{\small{Entropy density versus temperature in units of $T_1$. }}
%\label{fig:graph1}
\end{figure}
%______________________________________________________________

From this figure, we see that the entropy density is close to zero up to temperatures of order $ 0.4\,T_1\approx 50\,\text{MeV}$. Thus, in this 
temperature range, the entropy contributions are indeed negligible. It is obvious that the situation changes drastically with the temperature 
growth. Therefore, a natural question to ask is whether the entropy contributions play a major role at finite temperature. To answer this 
question, we consider the internal energy of the pair. As usual, it is given by ${\cal E}=F+T{\cal S}$. It is clear from above that ${\cal E}$ is 
independent of temperature at small distances, where it coincides with the heavy-quark potential at zero temperature. At large distances, the 
internal energy has a complex dependence on temperature together with a linear growth with $r$. Explicitly, it is given by 

\begin{equation}\label{large-E}
{\cal E}=\Sigma_{\text{\tiny T}}r
\,,
\end{equation}
with the tension 

\begin{equation}\label{Sigma}
\Sigma_{\text{\tiny T}}=\sigma_{\text{\tiny T}}\left(1+
\frac{8\sin^2\Bigl(\tfrac{1}{3}\arcsin\tfrac{T^2}{\TD^2}\Bigr)}{3-4\sin^2\Bigl(\tfrac{1}{3}\arcsin\tfrac{T^2}{\TD^2}\Bigr)}
\right)
\,.
\end{equation}

The difference with the string tension \eqref{sigma-T} is due to the second factor that is nothing but the entropy contribution. A direct but 
lengthy calculation shows that $\Sigma_{\text{\tiny T}}$ is a growing function of temperature. This is in contrast to the behavior of the string 
tension $\sigma_{\text{\tiny T}}$ discussed in section 3.1. Thus, the entropy contributions {\it do} play a major role at finite temperature. 

We conclude the discussion with a few short comments: 
\newline (i) A temperature growth of the singlet internal energy at large distances was observed in the lattice calculations of \cite{karsch2}. 
\newline (ii) The use of thermal AdS as the background metric in the low temperature phase results in zero entropy.\footnote{In this case 
$f\equiv 1$, so \eqref{ent} gives zero.} This leads to the picture that is completely inconsistent with physics of a pure $SU(3)$ gauge 
theory at finite temperature.\footnote{Note that the Hawking-Page transition between the thermal AdS space and the 
Schwarzschild black hole was discussed in \cite{HP} as a possible dual description of the deconfinement phase transition for large $N_c$ gauge 
theories.}
\newline (iii) In Fig.8 we plot $\Sigma_{\text{\tiny T}}$ versus temperature as provided by the expression \eqref{Sigma}.

%________________________  f - 8__________________________________
%
\vspace{.6cm}
\begin{figure}[htbp]
\centering
\includegraphics[width=5.5cm]{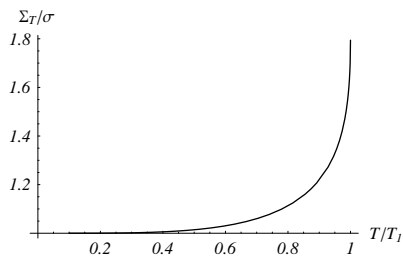}
\caption{\small{Tension $\Sigma_{\text{\tiny T}}$ in units of $\sigma$ versus temperature in units of $T_1$.}}
%\label{fig:graph1}
\end{figure}
%______________________________________________________________
\noindent (iv) It is tempting to see to what extent our predictions for the low temperature behaviors of $\sigma_{\text{\tiny T}}$ and 
$\Sigma_{\text{\tiny T}}$ are in agreement with the lattice data of \cite{karsch}. Unfortunately, the lattice data are only available for 
temperatures close to its critical value.\footnote{O.A. thanks O. Kaczmarek and P. Petreczky for a discussion of this issue.}
%_______________________________________________________________________________________
\subsection{Polyakov Loop}

As noted earlier, at very large separation the quark and anti-quark become free. In this case the dominant exponent is given by $S_\infty$. 
So, we are in a situation in which the approximation \eqref{approx} might be applicable. When we use it to find the expectation value of the 
Polyakov loop, we get 

\begin{equation}\label{LT}
L=\ep^{-\tfrac{1}{2}S_\infty}
\,,
\end{equation}
with $S_\infty$ the minimal area given by \eqref{S3}. Our normalization is stated in \eqref{cT} and \eqref{C}. With this choice, the Polyakov 
loop expectation value takes the form shown in Fig.9.

%________________________  f - 9__________________________________
%
\vspace{.6cm}
\begin{figure}[htbp]
\centering
\includegraphics[width=5.5cm]{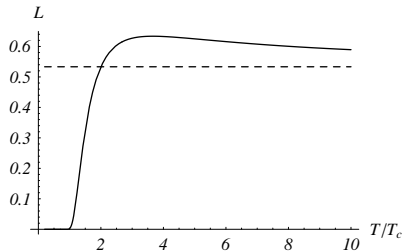}
\caption{\small{Polyakov loop expectation value versus temperature in units of $T_c$. The dashed line denotes the value $L_\infty$.}}
%\label{fig:graph1}
\end{figure}
%______________________________________________________________
Certainly, it has the desired behavior: The expectation value of $L$ is zero at low temperatures, while it is nonzero at high 
temperatures.\footnote{We consider an absolute value of $L$.} One sees there is a phase transition from the confined phase to the deconfined 
phase, as expected. Numerically, the transition temperature is of order

\begin{equation}\label{TcL}
T_c\approx 0.11\sqrt{c}\approx 100\,\text{MeV}
\,.
\end{equation}
This is roughly twice smaller than the value found in \cite{az2} from the spatial string tension.  Thus, the numerical consistency is not good 
enough at this point.

We chose the normalization constant in the form $c(T)=\alpha\,T+C$. The meaning of the coefficients is the following: $\alpha$ specifies the 
value of $L$ at $T=\infty$

\begin{equation}\label{Linfty}
L_\infty=\ep^{\,\tfrac{1}{2}(\g -\alpha)}
\,.
\end{equation}
$C$ specifies the form of $L$. Indeed, the expectation value of the loop is a continuously growing function of temperature for $C\geq0$, 
while it has a local maximum for $C<0$. The position of the maximum is given by a solution to equation

\begin{equation}\label{maxT}
CT+\frac{\g c}{\pi^2 }\int_0^1 dx\,\exp\{\,\tfrac{1}{2}\tfrac{c\,x^2}{\pi^2 T^2}\} =0
\,.
\end{equation}
For our set of parameters a numerical analysis of \eqref{maxT} results in $T\approx 3.8\,T_c$.

 Finally, we can compare the temperature dependence of the Polyakov loop expectation value as provided by our model with the lattice results of 
\cite{karsch}. Here there is a subtle point. The lattice data are only available for the range $1.03\leq \tfrac{T}{T_c}\leq 6$. This makes it difficult 
to see what exactly happens in the low temperature phase as well as for high temperatures. We now fix the normalization by fitting \eqref{LT} to 
the data given in Table 1 of \cite{karsch} near $T=6\,T_c$. In Fig.10 we have plotted $L$ against $\tfrac{T}{T_c}$. We find that the 
temperature dependence is in good agreement for $T\gtrsim 2\,T_c$.\footnote{Note that a better agreement is obtained by shifting the 
plot of Figure 10 a little bit to the left. }

%________________________  f - 10 __________________________________
%
\vspace{.6cm}
\begin{figure}[htbp]
\centering
\includegraphics[width=5.5cm]{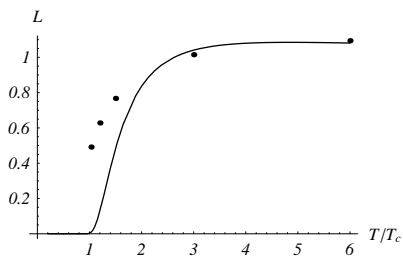}
\caption{\small{Polyakov loop expectation value versus temperature. Here $c(T)=0.96\,T-0.18$. The dots denote the data 
from \cite{karsch}.}}
%\label{fig:graph1}
\end{figure}
%______________________________________________________________
%________________________________________________________________________________________

\section{Discussion}
What we have learned is that the 5-dimensional effective model we proposed to study the properties of a heavy quark-antiquark pair at finite 
temperature turns out to be remarkably consistent with the qualitative expectations of thermal gauge theory. Moreover, in certain cases it 
provides the analytic results which are in good agreement with the lattice data. It is also worth noting that our model predicts the behavior of 
the thermodynamic quantities for low temperatures where field theory is unreliable and lattice data are missing. 

There are many issues that deserve to be further clarified. Let us mention some of them that are seemed the most important to us.

Apparently, the model suffers from a lack of numerical self-consistency: the value of the critical temperature as seen from the spatial 
string tension \cite{az2} turns out to be at least twice bigger than the value we found from the analysis of the Polyakov loop in section 3.3. 
There are two lines of thought on this problem. 

The first is to account for one-loop corrections in the world-sheet path integral \eqref{wilson}.\footnote{The model under consideration is 
an effective theory. It already includes some, but not all, quantum corrections in the approximation we used. The remaining corrections are 
due to string fluctuations.} This is a complicated 
problem. Indeed, it is a challenge to theorists to find the world-sheet formulation of string theory on warped 
geometries like AdS spaces. Among many things, it requires world-sheet fermions and even a world-sheet theta angle that is a two-form 
field $B$ with an arbitrary value of $\int B$ \cite{B}. But if it was resolved, it could help us to sum the series \eqref{wilson2} and hence
refine the estimates. There is another interesting problem here. What were considered in \cite{T-models} are some low temperature 
corrections to the string tension. It will be interesting to find similar corrections in our model.

The second line of thought is to somehow modify the background. It may include a slight revision of the metric \eqref{metric} or more radical 
changes like additional background fields. For instance, a recent proposal of \cite{csaki} is even to include the tachyon background.
If a modification is made, it would produce a number of additional free fit parameters that makes it less attractive for phenomenology. To escape 
the problem, a clever mechanism for reducing the number of parameters must be invented.

Part of the interest of AdS/CFT(QCD) stems from attempts to understand the physics of RHIC. The reasons are the following: First, a 
quark-gluon plasma is strongly coupled, so the perturbative QCD is of limited utility. Second, the lattice data do not always provide good 
information like, for instance on transport properties. On the other hand, AdS/CFT(QCD) offers an alternative way of dealing with this 
real-world problem.\footnote{For a recent review, see \cite{gubser}.} It would be particularly interesting to see if our model can shed some light 
on this subject.

\vspace{.25cm}

{\bf Acknowledgments}

\vspace{.25cm}
O.A. would like to thank O. Kaczmarek, P. Petreczky, M.I. Polikarpov, and E. Shuryak for useful communications and conversations.
The  work  of O.A. was supported in part by Max-Planck-Gesellschaft, Deutsche Forschungsgemeinschaft, and Russian Basic Research 
Foundation Grant 05-02-16486. 
%__________________                      R E F S                    ______________________

%____________________________________________________________________

\small
\begin{thebibliography}{99}
\bibitem{az2}
O. Andreev and V.I. Zakharov, The Spatial String Tension, Thermal Phase Transition, and AdS/QCD, hep-ph/0607026.
\bibitem{oa}
O. Andreev, Phys.Rev.D 73, 107901 (2006).
\bibitem{svet}
L.D. McLerran and B. Svetitsky, Phys. Lett.B 98, 195 (1981).
\bibitem{karsch2}
O. Kaczmarek, F. Karsch, P. Petreczky, and F. Zantow, Nucl.Phys. Proc.Suppl. 129, 560 (2004).
\bibitem{Tloops}
The literature on the Wilson loops at finite temperature is vast. The following is an incomplete list: \\
E. Witten, Adv.Theor.Math.Phys. 2, 505 (1998);\\
S.-J. Rey, S. Theisen, and J.-T. Yee,  Nucl.Phys.B 527, 171 (1998);\\
A. Brandhuber, N. Itzhaki, J. Sonnenschein, and S. Yankielowicz, JHEP 9806, 001 (1998);\\
D.J. Gross and H. Ooguri, Phys.Rev.D 58, 106002 (1998);\\
H. Dorn and H.J. Otto, JHEP 9809, 021 (1998);\\
 S.A. Hartnoll and S. P. Kumar, Phys.Rev.D 74, 026001 (2006);\\
S.-J. Sin and I. Zahed, Ampere's Law and Energy Loss in AdS/CFT Duality, hep-ph/0606049;\\
H. Boschi-Filho, N.R.F. Braga, and C.N. Ferreira, Phys.Rev.D 74, 086001 (2006);\\
K. Kajantie, T. Tahkokallio, and J.-T. Yee, Thermodynamics of AdS/QCD, hep-ph/0609254;\\
H. Boschi-Filho and N.R.F. Braga, AdS/CFT correspondence and strong interactions, hep-th/0610135.
\bibitem{effpot}
D.A. Kirzhnits and A.D. Linde, Phys.Lett.B 42, 471 (1972);\\
L. Dolan and R. Jackiw, Phys.Rev.D 9, 3320 (1974);\\
S. Weinberg, Phys.Rev.D 9, 3357 (1974).
\bibitem{karsch}
O. Kaczmarek, F. Karsch, P. Petreczky, and F. Zantow, Phys.Lett.B 543, 41 (2002).
\bibitem{az}
O. Andreev and V.I. Zakharov, Phys.Rev.D 74, 025023 (2006).
\bibitem{cornell}
E. Eichten, K. Gottfried, T. Konoshita, K.D. Lane, and T.-M. Yan, Phys.Rev.D 17, 3090 (1978); 
21, 203 (1980).
\bibitem{gr}
I.S. Gradshteyn and I.M. Ryzhik, Table of Integrals, Series, and Products, Academic Press, 1994. 
\bibitem{HP}
E. Witten, as cited in \cite{Tloops};\\
C.P. Herzog,  A Holographic Prediction of the Deconfinement Temperature, hep-th/0608151.
\bibitem{B}
See the discussion of E. Witten in \cite{Tloops}.
\bibitem{T-models}
R.D. Pisarski and O. Alvarez, Phys.Rev.D 26, 3735 (1982).
\bibitem{csaki}
C. Csaki and M. Reece, Toward a Systematic Holographic QCD: A Braneless Approach, hep-ph/0608266.
\bibitem{gubser}
S. Gubser, Relativistic heavy ion collisions and string theory, lectures given at the PiTP 2006 program, \\
http://www.admin.ias.edu/pitp/2006files/Lecture notes/Gubser Lecture notes REVISED.pdf. 
\end{thebibliography}
\end{document}